\documentclass[11pt,a4paper]{article}

\usepackage[dvips]{graphicx}
\usepackage[normal,small,bf]{caption2} 

\def\bfig#1{\begin{figure}[#1] \setcaptionwidth{12cm} \renewcommand{\captionlabeldelim}{.\ } \centering  \addtolength{\fboxsep}{5mm} }
\def\efig{\end{figure} \pagebreak[1]}

\newcommand{\be}{\begin{equation}}
\newcommand{\cf}{{\it cf. }}

\newcommand{\Ckk}[1]{C_{k'}^{k}(#1)}
\newcommand{\Ckkk}[1]{C_{k',-k'}^{k}(#1)}
\newcommand{\diff}{{\mathrm d}}
\newcommand{\ee}{\end{equation}}
\newcommand{\Ef}{E_{\mathrm{f}}}
\newcommand{\eg}{{\it e.g.}}
\newcommand{\Ei}{E_{\mathrm{i}}}
\newcommand{\equa}[1]{eq.~(\ref{#1})}

\newcommand{\ie}{{\it i.e.}~}

\newcommand{\sk}{\sqrt{\kappa_0^2-k^2}}

\newcommand{\xii}{x_{\mathrm{i}}}

\begin{document}

\title{A quantum evaporation effect}
\author{D. Boos\'e\footnote{E-mail: boose@lpt1.u-strasbg.fr.
Laboratoire de Physique Th\'eorique is
UMR 7085 of Centre National de la Recherche Scientifique and 
Universit\'e Louis Pasteur, Strasbourg.}
 and F. Bardou\footnote{E-mail: Francois.Bardou@ipcms.u-strasbg.fr.
 Institut de Physique et Chimie des Mat\'eriaux de Strasbourg 
 is UMR 7504 of Centre National de la Recherche Scientifique and
 Universit\'e Louis Pasteur, Strasbourg.}}
\date{published in {\it Europhys. Lett.} {\bf 53} (2001) pp. 1-7}

\maketitle
      
\begin{abstract}
A small momentum transfer to a particle interacting with a steep potential
barrier gives rise to a {\it quantum evaporation} effect which  
increases the transmission appreciably.
This effect results from the unexpectedly large population of quantum states 
with energies above the height of the barrier. Its characteristic 
properties are studied and an example of physical system in which 
it may be observed is given.
\end{abstract}

It is well known that the wave nature of quantum motion can amplify
as well as reduce quantum transport in comparison with its classical
counterpart. For example, a quantum particle is able to tunnel through
a potential barrier, a behaviour which is of course not possible in 
classical mechanics. On the contrary, the same 
particle is very likely to move in a certain part only of a  
random medium \cite{And58} whereas a classical particle may    
wander through the whole of it. The Schr\"odinger equation leads 
therefore to a large variety of situations regarding transport, whose study 
is still the subject of active 
research (see, \eg,~\cite{Spo96,MaS97,KKK97}).        
 
In this Letter, we describe a novel effect, called {\it quantum evaporation} 
hereafter, in which the wave nature of quantum motion amplifies  
transport appreciably. We study the behaviour in one dimension 
of a particle which undergoes a small momentum transfer while interacting 
with a rectangular potential barrier or with a potential step, both  
of height larger than its kinetic energy. We first present wave packet 
simulations of this behaviour, which reveal that a small momentum transfer 
is able to produce a large increase of the transmission into the 
classically forbidden region. We then explain this increase by relating it 
to the population, induced by the momentum transfer, of the quantum states 
with energies above the height of the potential. The population of these 
quantum states enables the particle to move in the classically inaccessible 
region, and so gives rise to quantum evaporation. We finally give an 
example of physical system in which this effect could be observed.

\bfig{!h}
\includegraphics[height=10.5cm,width=11.55cm,angle=0]{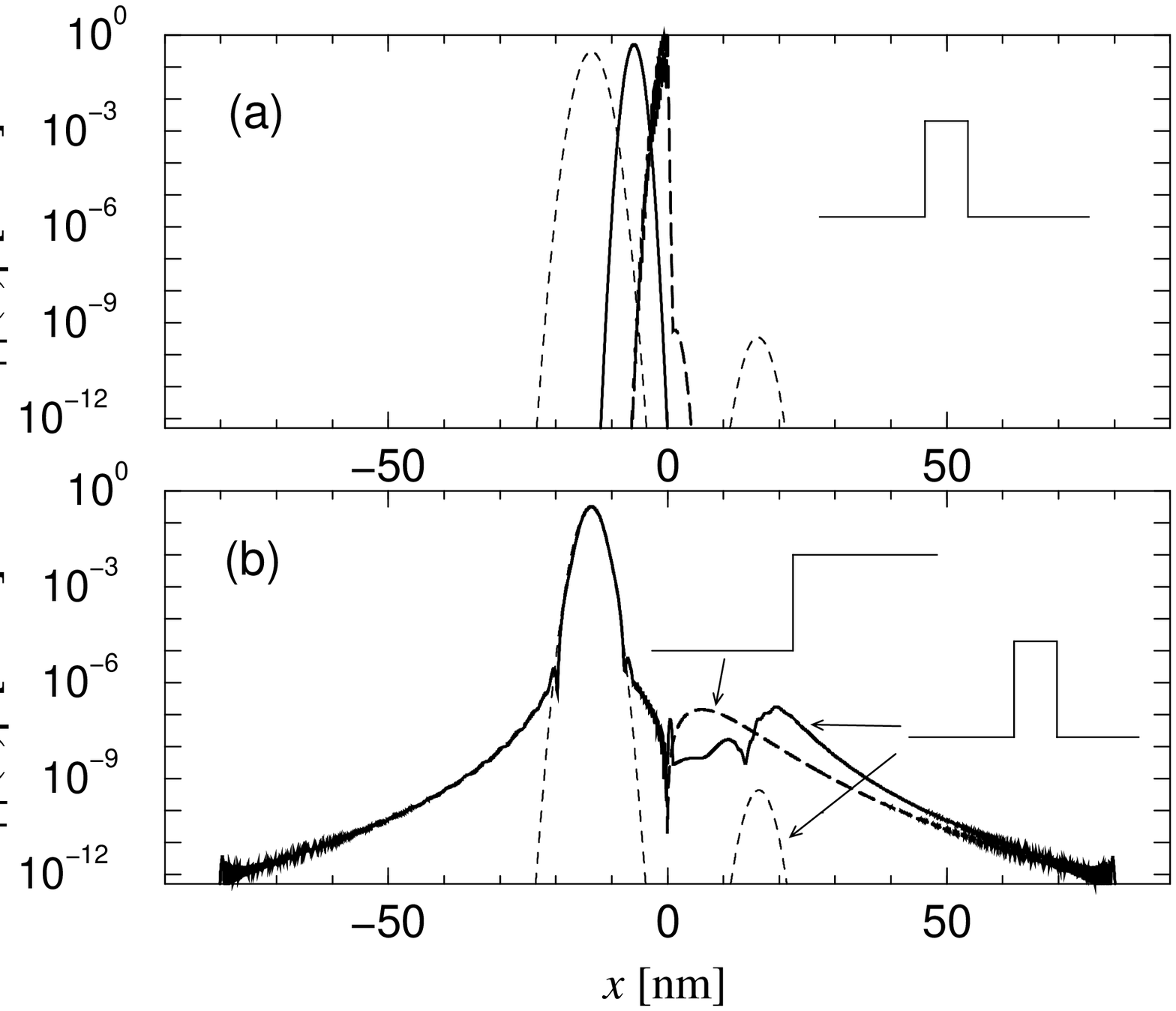}
\caption{Effect of a momentum transfer on the propagation of a
wave packet $\psi(x)$. a) {\it Without momentum
transfer}. At $t=0$~s, a wave packet of gaussian shape is sent
towards a potential barrier (solid line).
At $t\simeq 4.5\times 10^{-15}$~s,
the centroid of the wave packet reaches the potential barrier
(long-dashed line). At
$t\simeq 15\times 10^{-15}$~s, the transmitted and reflected wave packets are
well separated (dashed line).
b) {\it With a momentum transfer}. The shapes of the reflected and transmitted
wave packets are plotted at $t\simeq
15\times 10^{-15}$~s. Dashed line: transfer at $t=0$~s
and potential barrier. Solid line : transfer at
$t\simeq 4.5\times 10^{-15}$~s and potential barrier.
Long-dashed line : transfer at $t\simeq 4.5\times 10^{-15}$~s
and potential step. See text for details.
}
\label{fig1}
\end{figure}
    
Fig.~\ref{fig1} shows the shape of a quasi-monochromatic
wave packet at a few given times of its interaction
with a potential
\footnote{The evolution of the wave packet is obtained by
numerical integration of the time-dependent Schr\"odinger equation
according to the standard Crank-Nicholson method~\cite{ref_num}.
Grid spacings of $2.5\times10^{-18}$~s in time and 
$1.5\times 10^{-12}$~m in space have been used in our calculations.
The resulting accuracy on the values of  
the transmission probability $T$ is
better than 3\%.
Note that the results of our wave packet simulations 
are quite general because they can be adapted 
to any quantum particle with the help of
the usual scaling relations of the Schr\"odinger equation for time,
length and wave number.
}. 
This
wave packet has an initial shape which is gaussian, with the centroid at 
the position $\xii = -6$~nm and the standard deviation $\sigma = 0.8$~nm.
Its initial wave number
distribution is centred on the average wave number
$\overline {k}\simeq 1.2\times10^{10}\;\mathrm{m}^{-1}$ and has   
the standard deviation $\delta k = 1/(2{\sigma})\simeq 0.05 \overline {k}$. 
In addition, the wave packet has an average initial kinetic energy 
$\Ei=\hbar^2 \overline {k^2} /(2m)=5$~eV ($m$ is the mass of the particle),
a typical energy for an electron in a metal.
In fig.~\ref{fig1}a, it interacts with a rectangular potential barrier
of height $V_0=10$~eV which extends from $x=0$ to $x=1$~nm, a typical
barrier in solid state physics. The resulting transmission probability, 
$T \simeq 1.2\times 10^{-9}$, is of the same order of magnitude 
as the transmission probability $T \simeq 4.5\times 10^{-10}$ for a
purely monochromatic wave. It is much larger than the classical  
probability to have an energy larger than $V_0$, which is only 
$0.5\ \textrm{erfc}\left((\sqrt{mV_0}/\hbar - \overline {k}/\sqrt{2})/\delta k 
\right) \simeq 0.7 \times 10^{-15}$. 

In fig.~\ref{fig1}b, the wave packet undergoes a small instantaneous
momentum transfer 
$\hbar q$, with $q = 10^8$~m$^{-1}\simeq 10^{-2} \overline {k}$. 
If the transfer 
occurs at time $t=0$~s (dashed line), \ie much before the time  
$t_0 = m |x_i|/(\hbar \overline {k})\simeq 4.5\times 10^{-15}$~s
at which the centroid of the wave packet reaches the potential, the 
transmission probability increases only slightly, reaching the value 
$T \simeq 1.5\times 10^{-9}$. This is in agreement with the related 
small increase of the average kinetic energy from $\Ei=5$~eV 
to $\Ef=\hbar^2 \overline {(k+q)^2}/(2m)\simeq 5.09$~eV. On the  
contrary, if the transfer happens at a time very close to 
$t_0$ (solid line),  
the transmission probability increases up to 
$T \simeq 1.1\times 10^{-6}$. Such a large increase of the transmission 
probability, whose study is the subject of this Letter, can no more 
be explained by the variation of the average kinetic energy.
The shape of the transmitted
wave packet, plotted at time $t \simeq 15\times 10^{-15}$~s, is 
no longer gaussian whereas it is still nearly so in the case of
the transfer at $t=0$~s.

Fig.~\ref{fig1}b shows also the shape of the wave packet in the 
case of a potential step $V(x)=V_0H(x)$ ($H(x)$ is the Heaviside function)
and of the momentum transfer at a time $t \simeq t_0$
(long-dashed line). In the region $x>1$~nm, the transmitted wave packet 
is unexpectedly similar to the one in the case of the 
potential barrier. As time progresses,  
it propagates in the classically forbidden region $x>0$,
a behaviour which would hardly be possible 
with a momentum transfer taking place much before the time $t_0$.
The corresponding transmission probability,
$T \simeq 1.4\times 10^{-6}$, has nearly the same value as in the case     
of the potential barrier. These observations together with the preceding ones   
inevitably lead to the following conclusion.
A small momentum transfer which comes about  
while the centroid of the wave packet is close enough to the potential  
populates the quantum states with energies above $V_0$, even though 
the average final kinetic energy is less than $V_0$.     
The population of these states is then responsible 
for the large increase of the transmission probability observed in the
wave packet simulations. Thus, in spite of being  
of negligible weight in the wave packet before momentum transfer, the 
quantum states with energies above $V_0$ do play a crucial role
after the momentum transfer has happened.

\bfig{!h}
\includegraphics[height=12cm,width=9cm,angle=270]{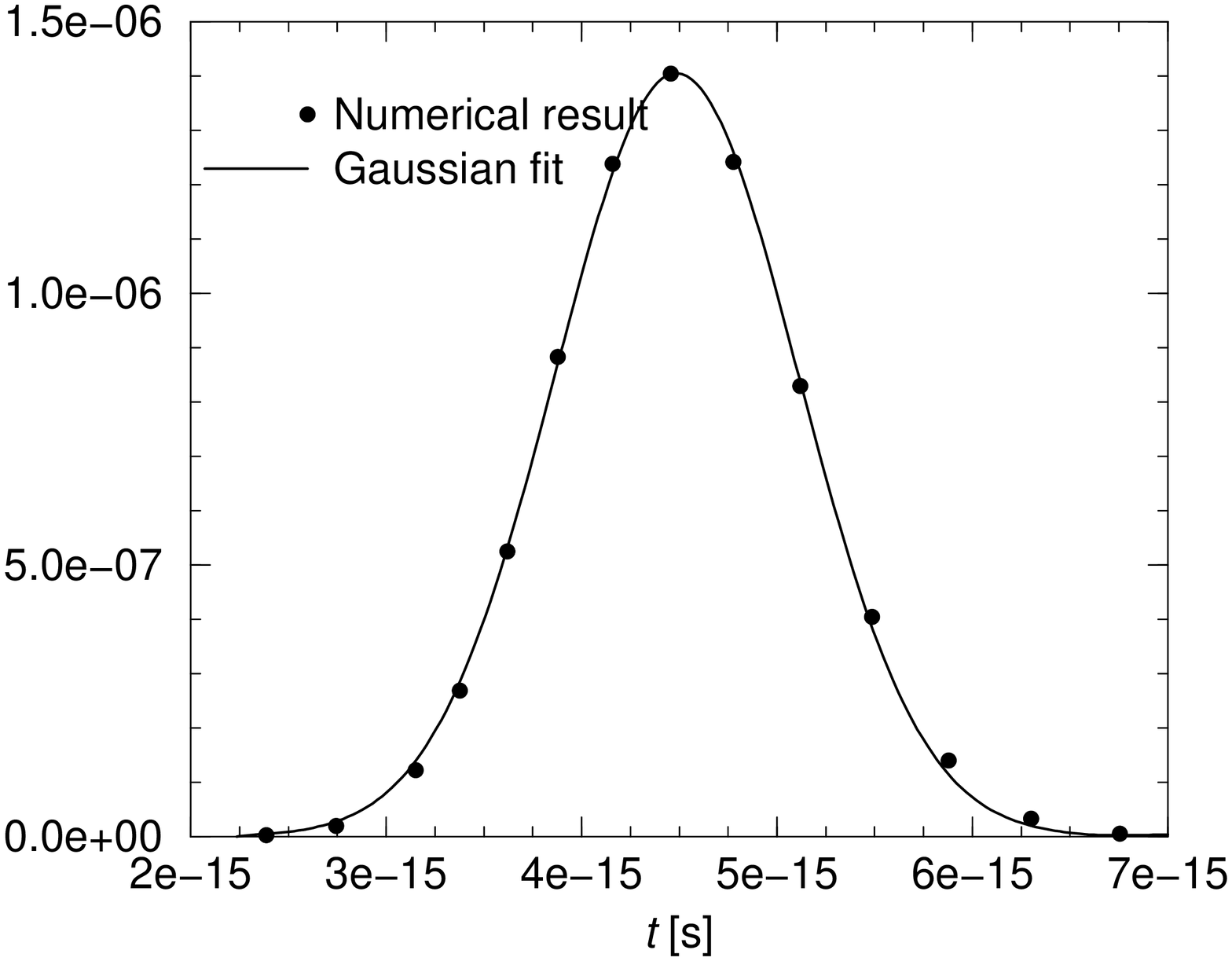}
\caption{Variation of the transmission probability $T$ with the time $t$ of
occurrence of the momentum transfer in the case of the potential
step. All parameters as in fig.~\ref{fig1}.
}
\label{fig2}
\efig

In order to identify the origin of the observed effect, we have examined
more precisely the influence of the {\it time} at which the momentum 
transfer takes place. Fig.~\ref{fig2} shows that the transmission probability
$T$ depends approximately as a Gaussian
$T_{\mathrm {max}} \exp\left(-(t-t_0)^2/2(\Delta t)^2\right)$ on the time $t$ 
of occurrence of the momentum transfer. 
The time at which the transmission probability takes its 
largest value $T_{\mathrm {max}} \simeq 1.4\times 10^{-6}$ 
is precisely the time 
$t_0\simeq 4.5\times 10^{-15}$s at which the centroid of the 
wave packet reaches the potential step. 
The length $2\Delta t\simeq 1.2\times 10^{-15}$s of the time interval
within which the momentum transfer must take place to produce a large 
increase of the transmission is found to be
nearly equal to the duration 
$2 m\sigma/(\hbar \overline {k})$ of appreciable interaction of the wave packet
with the potential. If the momentum transfer comes about at a time
$t\ll t_0 - \Delta t$, \ie much {\it before} the interaction of the 
wave packet with the potential, it shifts the initial wave number
distribution without reshaping it (see fig.~\ref{fig1}b, 
dashed line). The quantum states with energies above $V_0$ 
are then scantly populated and the resulting transmission is small. 
On the contrary, if the momentum transfer happens within the interval
$t_0 - \Delta t\leq t\leq  t_0 + \Delta t$, \ie {\it during} the  
interaction of the wave packet with the potential, it modifies largely  
the initial wave number distribution (see fig.~\ref{fig1}b,
solid and long-dashed lines). The quantum states with energies above $V_0$
are then well populated and the resulting transmission is large.

We have also studied the influence of the {\it duration} 
$\delta t$ of the momentum transfer.
A non-instantaneous transfer is found to produce the same increase
of the transmission
as an instantaneous one, provided that it
is fast enough (\ie $\delta t \leq 10^{-16}$~s). 
Thus, we focus only on instantaneous momentum transfers in the sequel.
It should be noted that the observed effect is {\it not} interpretable as a 
trivial consequence of a (naively applied) time-energy uncertainty relation.
Indeed, the energy spread $\delta E = \hbar/\delta t$ which is supposed 
to correspond to $\delta t = 10^{-16}$~s would be of the order 
of $V_0$. It would therefore give rise to a transmission of order
unity, which is obviously incompatible with our results. As shown
below, the energy distribution after momentum transfer does definitely
not result from a time-energy uncertainty relation.

In order to find the characteristic properties of the observed 
increase of the transmission, we use the following model. Our
initial wave packet includes only eigenfunctions $\psi_k(x)$ of the 
Hamiltonian describing the motion of the particle in the presence of the 
potential which have an energy below $V_0$. In the case of the potential
step, the expression of these eigenfunctions is
($0\leq k<\kappa_0=\sqrt{2mV_0}/\hbar$)
\begin{eqnarray}
 \psi_k(x) & = & \frac{1}{\sqrt{2\pi}}
                       (1-H(x)) \left( e^{ikx} + \frac{k-i\sk}{k+i\sk} \;
                                 e^{-ikx} \right) \nonumber \\
       & + & \frac{1}{\sqrt{2\pi}} H(x) \frac{2k}{k+i\sk} 
                                        e^{-\sk x} \ \cdot
\label{etat_ini}
\end{eqnarray} 
We take here the potential step in preference to the potential barrier so as to 
emphasize the fact that the studied effect is genuinely different from  
tunnelling~\footnote{\label{note_abruptness}
The results of this Letter are in fact valid for any potential
whose slope on the side where the particle comes in varies 
rapidly over a de Broglie wavelength.
Generally speaking, the steeper the slope, the larger the
transmission due to quantum evaporation.}.
The transfer of a momentum $\hbar q$ is modelled by a quantum jump,
as in the stochastic wave function description of the evolution of 
open quantum systems \cite{DCM92}. This quantum jump induces a 
translation in momentum space of every eigenfunction in the wave packet, 
\ie $\psi_k(p) \rightarrow \psi_k(p-\hbar q) = \psi_{k,q}(p)$.
The expression of the wave function $\psi_{k,q}(x)$ 
which corresponds to a given eigenfunction $\psi_k(x)$ in the wave packet
and takes the effect of the momentum transfer into account is then 
readily obtained by a Fourier transformation of the translated 
wave function $\psi_k(p-\hbar q)$. It is  
\be
   \psi_{k,q}(x) = e^{iqx} \psi_k(x) \ \cdot
\label{etat_q}
\ee
We restrict ourselves 
to the case of small momentum transfers, characterized by the condition 
$\hbar^2 \overline {(k+q)^2}/(2m) < V_0$. 
This condition ensures of course that any transmission produced by a 
momentum transfer
cannot be due to a trivial increase of the kinetic energy above $V_0$.

The existence of a position-dependent phase factor on the right-hand side of 
\equa{etat_q} has the striking consequence that a wave function 
$\psi_{k,q}(x)$ is no more an eigenfunction of the Hamiltonian 
describing the motion of the particle in the presence of the potential. 
Hence, $\psi_{k,q}(x)$ may be expanded in terms of the eigenfunctions
of this Hamiltonian, \ie
\begin{equation}
 \psi_{k,q}(x) = \int_0^{\kappa_0} \diff k' \Ckk{q} \psi_{k'}(x) 
            + \int_{\kappa_0}^{+\infty} \diff k' 
               \left( C_{-k'}^k(q) \psi_{-k'}(x) 
               +  \Ckk{q} \psi_{k'}(x) \right)  \ ,
\label{etat_q2}
\end{equation}
with $\Ckkk{q}=\int_{-\infty}^{+\infty} \diff x
{\psi_{k',-k'}^{*}}(x) \psi_{k,q}(x)$. 
The first integral in this expansion includes 
the nondegenerate eigenfunctions with energies below $V_0$
(\cf \equa{etat_ini})
and the second one the doubly degenerate eigenfunctions 
with energies above $V_0$ ($|k'|>\kappa_0$). Because of the eigenfunctions
with energies above $V_0$, every wave function $\psi_{k,q}(x)$ has  
components which are free to 
propagate anywhere in space, thus giving rise to a finite probability of 
quantum evaporation into the classically inaccessible region $x>0$.
    
If $f(k)$ denotes the amplitude corresponding to the initial wave number
distribution, the time-dependent wave packet which takes the effect of    
the momentum transfer into account is 
$\int_0^{\kappa_0} \diff k f(k) \psi_{k,q}(x,t)$.
The probability $T(q)$ of transmission into the classically forbidden region  
is then defined through the relation      
$T(q)=\lim_{t\to +\infty} \int_0^{+\infty} \diff x
|\int_0^{\kappa_0} \diff k f(k) \psi_{k,q}(x,t)|^2$.
A close study of the wave packet simulations shows that
$T(q)$ is dominated by the contributions with 
energies above $V_0$. The transmission probability may therefore be computed
accurately with the help of the following formula: 
\begin{equation}
T(q) \sim \lim_{t\to +\infty} \int_0^{+\infty} \diff x 
         \left| \int_0^{\kappa_0} \diff k f(k) 
                   \int_{\kappa_0}^{+\infty} \diff k'
           \; e^{-\frac{i\hbar k'^2}{2m} t}   
         \left(  C_{-k'}^k(q) \psi_{-k'}(x) 
                  + \Ckk{q} \psi_{k'}(x) \right) \right|^2 \cdot
\label{T}
\end{equation}
A tedious yet straightforward calculation using the expressions of the  
wave functions $\psi_{k,q}(x)$ and of the 
eigenfunctions $\{ \psi_{k'}(x), \psi_{-k'}(x) \}$ with 
energies above $V_0$ leads to the following formulae for 
the amplitudes $C_{-k'}^k(q)$ and $C_{k'}^k(q)$ 
in \equa{T}:
\begin{eqnarray}           
C_{-k'}^k(q) = & &\frac{4i \kappa_0^2 k\sqrt{k'^2- \kappa_0^2 }}       
               {\pi \left(k'+\sqrt{k'^2- \kappa_0^2 }\right)
               \left(k+i\sqrt{ \kappa_0^2 -k^2}\right) } \nonumber \\
        &\times & \frac{q}{\left((k'+q)^2-k^2\right)}  \;    
        \frac{1}{\left(\sqrt{ \kappa_0^2 -k^2}-iq\right)^2+k'^2- \kappa_0^2 } \; ,  
\label{Ca}           
\end{eqnarray}
\begin{eqnarray}           
C_{k'}^k(q)  = & & \frac{4i \kappa_0^2 kk'}       
               {\pi \left(k'+\sqrt{k'^2- \kappa_0^2 }\right)
               \left(k+i\sqrt{ \kappa_0^2 -k^2}\right) } \nonumber \\
   & \times &  \frac{q}{\left(k'^2-(k+q)^2\right)\left(k'^2-(k-q)^2\right)} \;
     \frac{\left(\sqrt{ \kappa_0^2 -k^2} +i\sqrt{k'^2- \kappa_0^2 } +iq\right)}
     {\left(\sqrt{ \kappa_0^2 -k^2}+i\sqrt{k'^2- \kappa_0^2 }-iq\right)} \;\cdot
\label{Cb}           
\end{eqnarray}

A first notable property of quantum evaporation follows directly from the  
expressions of the amplitudes $C_{-k'}^k(q)$ and $C_{k'}^k(q)$. 
Eqs. (\ref{Ca}) and (\ref{Cb}) show that both amplitudes are ratios of 
products of 
algebraic functions of the wave numbers $k$, $k'$ and $q$. Consequently, the 
effect of quantum evaporation decreases only {\it algebraically} in $k'$ 
if this wave number increases from $\kappa_0$ up to infinity. This 
algebraic decrease explains why quantum evaporation produces much larger 
transmissions than does tunnelling (whose effect decreases {\it exponentially}
in $\sqrt{ \kappa_0^2 -k^2}$ if this wave number increases from zero up to
$\kappa_0$), as one can see by comparing the curves (b) and (c) to the 
curves (a) in fig.~\ref{fig3} below.  

Since quantum evaporation has a large effect on the  
transmission, it is interesting to examine the magnitude of the  
energy transfers. The average energy $\Ef$ of the particle  
after momentum transfer is~\cite{BaB99}
\be
\Ef = \hbar^2 (\overline {k^2} + q^2)/(2m) < V_0\ \cdot
\label{enfi}
\ee       
Thus, for small momentum transfers $\hbar q$ such that $|q| \ll \overline {k}$, the 
energy transfer $\Ef - \Ei = \hbar^2 q^2/(2m)$ is negligible in comparison with
the average initial energy $\Ei$. This point reveals another remarkable property 
of quantum evaporation, namely that an appreciable increase of the 
transmission does not require a large variation of the average 
energy. This is so because the multiplication of any  
$\psi_k(x)$ by a factor $e^{iqx}$ (\cf~\equa{etat_q}) removes  
the stationary character of this 
wave function, and so increases the transmission, 
irrespective of the magnitude of the transferred
wave number $q$. The smallness of the variation of the average
energy comes from the fact that the momentum transfer populates 
states with energies above as well as below $\Ei$. It should be noted    
that a small momentum transfer {\it before} the interaction of  
the wave packet with the potential produces an average energy transfer 
$\Ef-\Ei= \hbar^2(\overline {(k+q)^2}-\overline {k^2})/(2m) 
\simeq \hbar^2{\overline {k}}q/m$.
Interestingly enough, this average energy transfer is far larger than the 
previous one, even though the resulting transmission 
is much smaller than in the case of quantum evaporation.

The amplitudes $C_{-k'}^k(q)$, \equa{Ca}, and $C_{k'}^k(q)$, \equa{Cb}, are 
proportional to the transferred wave number $q$ in the regime 
$|q|\ll (\kappa_0^2-k^2)/(2\kappa_0)$~\cite{BaB99}. The transmission 
probability, \equa{T}, has then the following simple expression: 
\be
T(q) \propto q^2 \, \cdot
\label{T2}
\ee
Therefore, whether the particle undergoes a 
forward ($q>0$) or backward ($q<0$) momentum transfer, the resulting 
transmission increases by the same amount if 
$|q|\ll (\kappa_0^2-k^2)/(2\kappa_0)$.
This insensitivity to the sign of $q$  
is a third characteristic property 
of quantum evaporation. It can be understood by remembering  
that the multiplication of any $\psi_k(x)$ by a factor 
$e^{iqx}$ (\cf~\equa{etat_q}) removes the stationary character of this  
wave function, and so increases 
the transmission, irrespective of the sign of the transferred 
wave number $q$. By comparison, the effect of a momentum 
transfer coming about {\it before} the interaction of the wave packet with the 
potential amounts merely to a trivial shift in momentum, and  
the resulting transmission then increases or decreases according to 
whether $q$ is positive or negative.

\bfig{!h}
\includegraphics[height=12cm,width=9cm,angle=270]{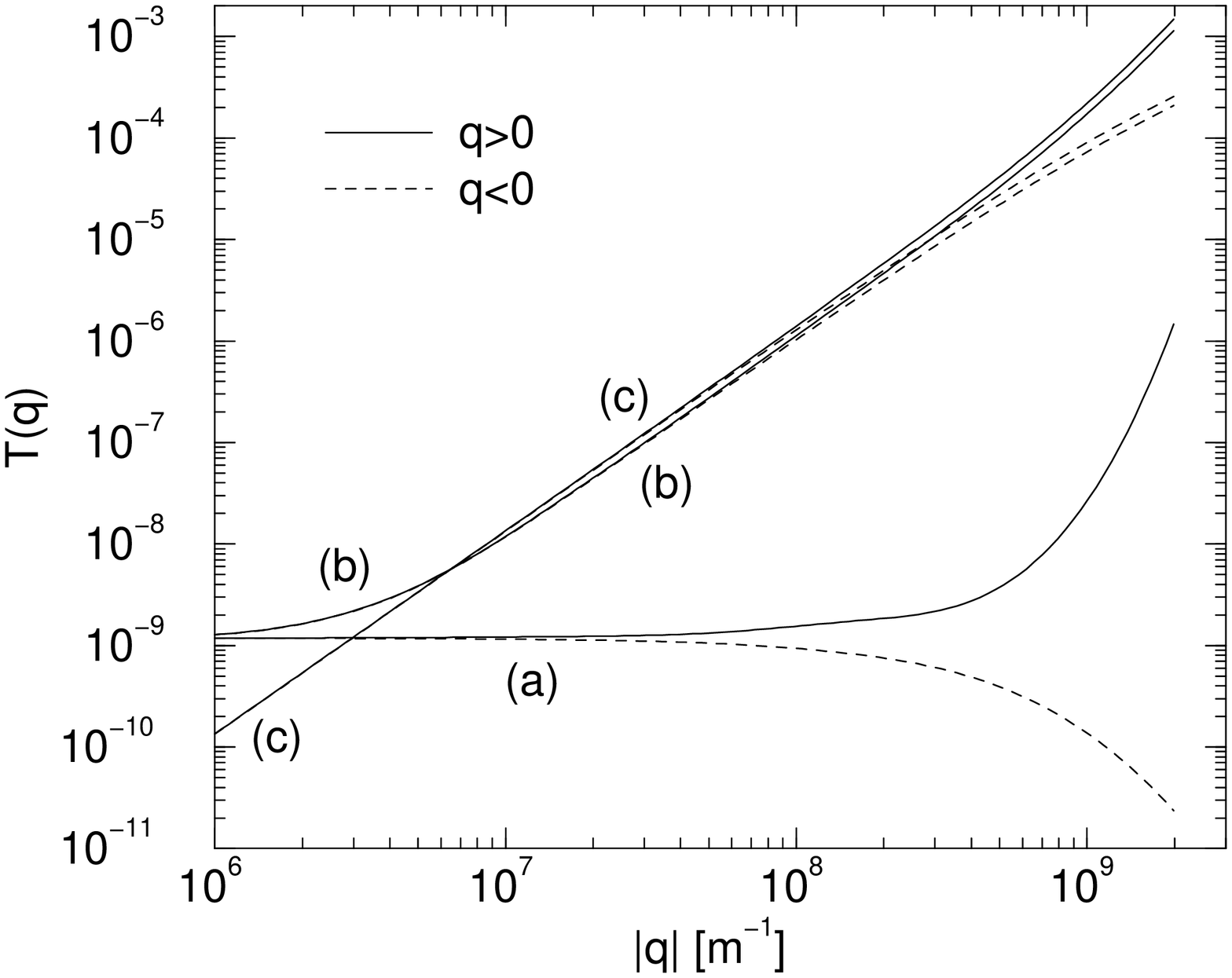}
\caption{Variation of the transmission probability $T$ with the
transferred wave number $q$.
Wave packet and potential parameters as in fig.~\ref{fig1}
($\overline {k}\simeq 1.2\times10^{10}$~m$^{-1}$).
a) Transfer {\it before} interaction with the potential barrier.
b) Transfer {\it during} the interaction with the potential barrier.
c) Transfer {\it during} the interaction with the potential step.
}
\label{fig3}
\efig

Fig.~\ref{fig3} shows the variation of the transmission
probability $T$ with the transferred wave number $q$
for the cases considered in fig.~\ref{fig1}b.  
In the case of a momentum transfer {\it before} the interaction
with the potential barrier (curves (a)), the variation of $T(q)$
is small and depends on the 
sign of $q$ because the transmission is mainly
due to tunnelling. On the contrary, in the case of a
momentum transfer {\it during} the interaction with the
potential barrier (curves (b)), $T(q)$  
increases by several orders of magnitude 
because the transmission is fully dominated by quantum evaporation.
For any value of $q$ 
in the regime $|q| \ll (\kappa_0^2-k^2)/(2\kappa_0)$, 
$T(q)$ is independent of the sign of $q$ and varies
quadratically (up to $|q|\simeq 0.02\overline {k}$), in 
agreement with \equa{T2}. The energy transfer becomes
of course important at larger values of $q$, with the expected consequence
that a forward momentum transfer produces a larger transmission 
than does a backward one. In the case of a momentum transfer {\it during}
the interaction with the potential step (curves (c)), the transmission is 
of course caused by quantum evaporation only. The corresponding curves 
are nearly identical to the curves (b), which confirms the fact 
that tunnelling has a negligible effect on the transmission in the case
of the potential barrier.   

Finally, we would like to point out that there are physical systems which 
may be used to detect the existence of quantum evaporation. For 
instance, systems 
in which electrons are field emitted from a metal surface upon 
the application of a strong electric field could be employed with this aim
in view. In such systems, quantum evaporation 
could result from electron-electron or electron-phonon scattering events 
taking place in the metallic tip. It would then lead to the appearance of a  
high energy tail with a telltale shape in the spectrum 
of field emitted electrons~\cite{note_tails}. 
Laser cooled atomic gases should prove to be even more interesting for the
observation of the  
effect. This is so because, by switching a resonant laser beam on and
off, one controls both the value and the time of 
occurrence of a momentum transfer to any such system.   
We have considered a numerical example in which cold metastable helium atoms 
are sent with an average initial kinetic energy of $10^{-11}$~eV
towards a potential step whose mean height is equal to  
$1.5\times 10^{-11}$~eV. Owing to recent advances in the field of laser 
cooling techniques, kinetic energies of such small values are now
reachable in practice~\cite{note5}. The corresponding
de Broglie wavelengths are then sufficiently large for a potential step
with a steep enough slope to be generated (\cf~note \ref{note_abruptness}).
Since the duration of interaction of a cold helium atom with the 
potential is of the order of $10^{-3}$~s, one has ample time to create a
momentum transfer by spontaneous emission of a photon. 
This can be done in practice by adjusting for instance the laser to 
the transition $2^3S_1 \to 2^3P_1$ (lifetime of $2^3P_1$ $\simeq 100$~ns).
Our calculations indicate that a backward wave number transfer of 
$q=-5\times10^{5}\;\mathrm{m}^{-1}$ (which corresponds to an energy transfer 
of $1.3\times 10^{-12}$~eV) increases the transmission 
probability from less than $10^{-10}$ up to $5\times10^{-4}$, thus
producing a potentially detectable effect.
Lastly, let us mention that one may also obtain a quantum evaporation 
effect by giving a velocity $v = \hbar q/m$ to the potential instead of 
transferring a momentum $\hbar q$ to the atoms, as can be shown by using a
Galilean transformation of the Schr\"odinger equation~\cite{BaB99}.

We thank A.~Aspect and G.~Ingold for stimulating discussions and the referees
for useful comments.

\pagebreak


\vskip-12pt
\begin{thebibliography}{99}
          
\bibitem{And58} P.W.~Anderson, Phys. Rev. {\bf 109} (1958) 1492.

\bibitem{Spo96} H.~Spohn, Phys. Rev. Lett. {\bf 77} (1996) 1198.

\bibitem{MaS97} M.S.~Marinov and B.~Segev, Phys. Rev. A {\bf 55} (1997) 3580.

\bibitem{KKK97} R.~Ketzmerick, K.~Kruse, S.~Kraut and T.~Geisel, Phys. Rev. Lett. {\bf 79} (1997) 1959.

\bibitem{ref_num} W.H.~Press, B.P.~Flannery, S.A.~Teukolsky and W.T.~Vetterling, {\it Numerical Recipes in C} (Cambridge University Press) (1991).

\bibitem{DCM92} For details, see, \eg, J.~Dalibard, Y.~Castin 
and K.~M\o lmer, Phys. Rev. Lett. {\bf 68} (1992) 580;
R.~Dum, P.~Zoller and H.~Ritsch, Phys. Rev. A {\bf 45} 
(1992) 4879 ; H.J.~Carmichael, {\it An Open Systems 
Approach to Quantum Optics} (Springer-Verlag, Berlin) 1993;
C.~Cohen-Tannoudji, F.~Bardou and A.~Aspect,
in {\it Proceedings of Laser Spectroscopy~X, Font-Romeu, 1991}, 
edited by M.~Ducloy, E.~Giacobino and G.~Camy
(World Scientific, Singapore) 1992, pp. 3-14.   

\bibitem{BaB99} F.~Bardou and D.~Boos\'e, in preparation.  

\bibitem{note_tails}
Such tails have been repeatedly observed (C.~Lea and R.~Gomer,
Phys. Rev. Lett. {\bf 25} (1970) 804; 
H.~Ogawa, N.~Arai, K.~Nagaoka, S.~Uchiyama, 
T.~Yamashita, H.~Itoh and C.~Oshima, Surf. Sci. {\bf 357-358}
(1996) 371), 
but a detailed theoretical description of them is still missing.  

\bibitem{note5} F.~Bardou, B.~Saubam\'ea, J.~Lawall, K.~Shimizu,
O.~Emile, C.~Westbrook, A.~Aspect and C.~Cohen-Tannoudji,
C. R. Acad. Sci. Paris S\'{e}rie II {\bf 318} (1994)
877; 
B.~Saubam\'ea, T.W.~Hijmans, S.~Kulin, E.~Rasel, E.~Peik, M.~Leduc
and C.~Cohen-Tannoudji, Phys. Rev. Lett. {\bf 79} (1997) 
3146. 
   

\end{thebibliography}
\end{document}